\documentstyle{article}

\newcommand{\r}{\rho}
\newcommand{\pp}{{1\over \hat{\epsilon}}}
\newcommand{\p}{-{1\over \hat{\epsilon}}}
\begin{document}
\begin{titlepage}
\begin{centering}
\title{Factorization in Semi-Inclusive Polarized  Deep Inelastic
Scattering\thanks{Partially
supported by CONICET-Argentina.}}
\author{D. de Florian, C.A.Garc\'{\i}a Canal \\
Laboratorio de F\'{\i}sica Te\'{o}rica \\
Departamento de F\'{\i}sica \\
Universidad Nacional de La Plata \\ C.C. 67 - 1900 La Plata -
Argentina \\ \\
R.Sassot  \\
Departamento de F\'{\i}sica \\
Universidad de Buenos Aires \\
Ciudad Universitaria, Pab.1 \\
(1428) Bs.As. -
Argentina}
\date{8 October 1995}
\maketitle
\end{centering}
\begin{abstract}
We calculate and analize the ${\cal{O}}(\alpha_s)$ one-particle inclusive cross
section
in polarized deep inelastic lepton-hadron scattering, using dimensional
regularization and the HVBM prescription for $\gamma_5$. We discuss the
factorization of all the collinear singularities related to the process,
particularly those which are absorbed in the redefinition of the spin dependent
analogue of the recently introduced fracture functions. This is
done in the usual $\overline{MS}$ scheme and in another one, called
$\overline{MS_p}$, which factorizes
soft contributions and guarantees the axial current (non)conservation
properties.
\end{abstract}

\noindent\hspace*{10mm} {\bf PACS}: 12.38.Bx, 13.85.Ni, 13.88.+e \\
\hspace*{10mm} {\bf Keywords}: Semi-Inclusive, Polarized, QCD corrections,
Fracture Functions
\end{titlepage}

\noindent{\large \bf Introduction}\\

In recent years, there has been an increasing theoretical and experimental
interest in semi-inclusive deep inelastic phenomena. Specifically, the use of
one-particle inclusive measurements, with polarized targets and beams, has been
indicated as an adecuate tool
to unveil the spin structure of the proton, elusive to the totally inclusive
experiments
(see \cite{hmc,hermes,rhic} and references therein).

However, the available calculations \cite{gull,frank,bass} of one-particle
inclusive polarized deep inelastic cross sections do not include the full QCD
next to leading order corrections, which are essential to weight the role of
the gluon
polarization \cite{scale}. These calculations are also not adequate for
phenomenological purposes because they are not able
to describe the full target fragmentation kinematical region \cite{pi}, that,
incidentally, is expected to be favored in the foreseen experiments \cite{hmc}.
Higher order corrections produce singularities in this region that are usually
avoided
imposing cuts in the transverse momentum allowed for the produced particles
\cite{pi,grau}.

In order to cope with the problem of the target fragmentation, a new
factorization
approach for semi-inclusive processes has been introduced by Trentadue and
Veneziano \cite{ven},
defining new unpertubative distributions, called fracture functions. These
distributions measure the probability for finding a parton and a hadron in the
target and can be measured in the proposed experiments. The use of
this approach in next to leading order one-particle inclusive unpolarized deep
inelastic scattering, has been recently shown \cite{grau} to allow a consistent
factorization of the
collinear singularities coming from the kinematical region where the hadron is
produced in the direction of the incoming nucleon,  and which cannot be
absorbed in the
redefinition of the usual distributions.

The extension of this approach to polarized phenomena using dimensional
regularization
\cite{jj}, implies an arbitrariness regarding the definition used for the
$\gamma_5$ matrix. Between the different prescriptions, the one proposed in
reference  \cite{hvbm}  (HVBM),
has been proved to be fully consistent and extensible to any order in
perturbation theory.
However, this prescription introduces finite soft contributions that come from
the
breaking of chiral invariance and have to be substracted in the distribution
functions, withdrawing from the $\overline{MS}$ scheme \cite{vog,heavy}. It is
important, then, to
show explicitely that the substraction rule used for polarized parton
distributions in totally
inclusive processes factorizes the same singularities and soft terms in those
which are one-particle inclusive and can be generalised
for fracture functions in a completely consistent way.

In the following section we define the spin dependent one-particle inclusive
cross section in terms of the polarized structure and fracture functions and
the
unpolarized fragmentation function. In the third we show the results for the
unsubstracted ${\cal{O}}(\alpha_s)$ contributions coming from the relevant
diagrams using the HVBM prescription. Finally, we discuss the factorization of
collinear singularities and the rules for the substraction of finite soft terms
in the different
factorization schemes.\\

\noindent{\large \bf Definitions and kinematics.}\\

In this section we introduce the spin dependent fracture function, generalizing
what has been done in references \cite{grau,ven}, and we
establish our notation.

In the one photon exchange approximation for the interaction between a lepton
of momentum $l$ and helicity $\lambda$ and a nucleon $N$ of momentum $P$ and
helicity $\lambda '$, the differential cross section for the production of $n$
partons can be written as:
\begin{eqnarray}
\frac{d \sigma^{\lambda \lambda '}}{dx\,dy\,dPS^{(n)}}= \hspace*{130mm}  \\
\sum_{i=q,\bar q , g}\sum _{\lambda ''=\pm 1}\int \frac{d \xi}{\xi} \,
P_{i/N}(\xi,\frac{\lambda ''}{\lambda '}) \,
\frac{\alpha^2}{S_{H}x}\frac{1}{e^2
(2\pi)^{2d}}\left[
Y_M(-g^{\mu\nu})+Y_L \frac{4x^2}{Q^2}P_\mu P_\nu + \lambda  Y_{P}\,
\frac{x}{Q^2}\, i \epsilon^{\mu \nu q P } \right] H_{\mu\nu}(\lambda '')
\nonumber
\end{eqnarray}
where
\begin{equation}
x=\frac{Q^2}{2 P\cdot q} , \ \ \ y=\frac{P\cdot q}{P\cdot l} ,\ \ \ S_H=(P+l)^2
\end{equation}
 being $q$ the transfered momentum ($Q^2=-q^2$) and
 $dPS^{(n)}$  the phase space of $n$  final state partons in $d=4-2 \epsilon$
dimensions.
$P_{i/N}(\xi,\lambda ''/\lambda ')$ is the probability for finding a parton $i$
with
helicity $\lambda ''$ carrying a fraction $\xi$ of the nucleon momentum, and
the kinematical factors appearing in the leptonic tensor are
\begin{eqnarray}
 Y_M=\frac{1+(1-y)^2}{2y^2} \  \ \ \ \ Y_L=\frac{4(1-y)+(1-y)^2}{2y^2}  \ \ \ \
\  Y_P=\frac{2-y}{y}
\end{eqnarray}
The helicity dependent partonic tensor  is defined by
\begin{eqnarray}
H_{\mu\nu}(\lambda'')=M_{\mu}(\lambda '')\, M^{\dagger}_{\nu}(\lambda '')
\end{eqnarray}
where $M_{\mu}$ is the parton-photon matrix element, with the photon
polarization vector
factorized out.

In order to isolate the antisymmetric part of this tensor, which leads to the
polarized structure function, we take the difference between cross sections
with opposite target helicities
\begin{eqnarray}
\Delta \sigma \equiv \sigma^{\lambda +}-\sigma^{\lambda -}
\end{eqnarray}
at variance with the unpolarized case where an average over beam and target
helicities is taken.
With these definitions
\begin{eqnarray}
\frac{d \Delta \sigma}{dx\,dy\,dPS^{(n)}}= \sum_{i=q,\bar q , g}\int \frac{d
\xi}{\xi}\, \Delta P_{i/N}(\xi)\,  \frac{\alpha^2}{S_{H}x}\, \frac{1}{e^2
(2\pi)^{2d}}\, \lambda  Y_{P}\, \frac{x}{Q^2} \,
i \epsilon^{\mu \nu q P}  \Delta H_{\mu\nu}
\end{eqnarray}
where
\begin{eqnarray}
\Delta H_{\mu\nu}\equiv M_{\mu}(+) M^{\dagger}_{\nu}(+)-
M_{\mu}(-) M^{\dagger}_{\nu}(-)
\end{eqnarray}
\begin{eqnarray}
\Delta P_{i/N}(\xi)\equiv P_{i/N}(\xi,+)- P_{i/N}(\xi,-)
\end{eqnarray}
In analogy with the unpolarized case, treted in reference  \cite{grau}, we
write the cross section for the production
of unpolarized hadrons h of energy $E_h$ with polarized beams and targets,
differential in the variable $z=E_h/E_N(1-x)$ as
\begin{eqnarray}
\frac{d \Delta \sigma}{dx\,dy\,dz}=  \int \frac{du}{u}\sum_N\sum_{j=q,\bar q ,
g} \int dPS^{(n)}
\,  \frac{\alpha^2}{S_{H}x}\, \frac{1}{e^2 (2\pi)^{2d}}\, \lambda Y_P
 \frac{x}{Q^2}\, i \epsilon^{\mu \nu q P} \Delta H_{\mu  \nu} \nonumber \\
\left\{ \Delta M_{j,h/A} \left( \frac{x}{u},\frac{E_h}{E_A}\right)(1-x)+\Delta
f_{j/A}\left(\frac{x}{u}\right)
\sum_{i_\alpha =q,\bar q , g }
D_{h/i_{\alpha}}\left(\frac{E_h}{E_{\alpha}}\right)
\frac{E_h}{E_{\alpha}}(1-x)\right\}
\end{eqnarray}
The variable $u$ is given, as usually,
by $u=x/\xi$. The spin dependent fracture function $\Delta
M_{j,h/N}(\xi,\zeta)$ is the probability for  finding a polarized parton $j$
with momentum fraction $\xi$ and a hadron h with momentum fraction $\zeta$ in
the nucleon N. Both $\Delta f_{j/N}$ and $D_{h/i_{\alpha}}$ are the usual spin
dependent parton distribution and fragmentation function respectively
\cite{ell}.
Notice that in the case of hadrons with spin, the fragmentation function is
exactly the unpolarized one due to the fact that we are summing over the final
state polarizations because they are not observed in the present experiment.
For spinless hadrons, this is also true provided the fragmentation mechanism is
independent of the helicity of the parent parton, as it is usually assumed
\cite{gull,frank}.\\

\noindent{\large \bf ${\cal{O}}(\alpha_s)$ contributions.}\\

In the following we calculate the spin dependent
cross section  up to order $\alpha_s$. For this purpose, it is convenient to
use the same kinematical variables as in the unpolarized case but contracting
the matrix elements $\Delta H_{\mu\nu}$ of the relevant processes with the
following projector
\begin{eqnarray}
P_{pol}^{\mu\nu}\equiv  \frac{\alpha^2}{S_{H}x}\,\, \frac{1}{e^2
(2\pi)^{2d}}\,\, \frac{x}{Q^2} \,i \epsilon^{\mu \nu q P}
\end{eqnarray}
This projector picks up  at tree level (Figure 1a) and after integrating over
the phase space for one particle, only contributions proportional to delta
functions in the convolution variables, being the proportionality factor
\begin{eqnarray}
c_j= 4\pi  Q^2_{q_j} \, {{\alpha^2}\over {S_H x}}
\end{eqnarray}
Then,
\begin{equation}
\frac{d\Delta \sigma}{dx\, dy\, dz}=\lambda Y_P \sum_{j=q,\bar{q}}c_j \left\{
M_{j}(x,(1-x)z)\, (1-x)+ \Delta f_{j}(x)\, D_{j}(z) \right\}
\end{equation}
where we have dropped the indeces labelling the target and produced hadron.
The virtual corrections (Figures 1b, 1c, 1d) give the same contribution but now
multiplied by the usual factor \cite{aem}
\begin{eqnarray}
\frac{\alpha_s}{2\pi}\left(\frac{4\pi \mu^2}{Q^2}
\right)\frac{\Gamma(1-\epsilon)}{\Gamma(1-2\epsilon)}C_f
\left(-2\frac{1}{\epsilon ^2}- 3\frac{1}{\epsilon}-8-\frac{\pi^2}{3} \right)
\end{eqnarray}
The results for  the real gluon emission (Figure 2)  and the box diagrams
(Figure 3) were calculated using the program {\sc Tracer} \cite{tracer} and can
be found in
apendix A. Notice that, as we are working in the HVBM scheme, terms
proportional
to the square of the $d-4$ dimensional component of the momentum of the
outgoing particles, $\hat p_{{\rm out}}^2$, must be isolated \cite{vog}.
Working in the photon-parton center of mass frame, there is no need to
discriminate between the two
outgoing particles, because they have opposite momenta. Furthermore, in this
frame the incoming particles do not have $d-4$ components of the momentum.
For fragmentation like configurations,  the two-particle phase space in which
the matrix elements are integrated,
is given by
\begin{eqnarray}
\label{dps}
dPS^{(2)} = \frac{1}{8\pi} \frac{(4\pi)^\epsilon}{\Gamma(-\epsilon)}
\frac{u-x}{u(1-x)}\, d\rho \int_0^{\hat{p}^2_{{\rm max}}} d\hat p_{{\rm out}}^2
\left(\hat p_{{\rm out}}^2\right)^{-1-\epsilon}
\end{eqnarray}
with
\begin{eqnarray}
 \hat{p}^2_{{\rm max}}= Q^2 (1-u)\, u\, (1-\rho) \left( \rho
-\frac{x(1-u)}{u(1-x)} \right) \left(\frac{1-x}{u-x}\right)^2 ,
\end{eqnarray}
where the variable $\rho$ is defined by
\begin{equation}
\rho\equiv\frac{E_\alpha}{E_N (1-x)}
\end{equation}
i.e., the energy fraction of the parton $\alpha$ which undergoes hadronization.
Due to the fact that all the contributions can be decomposed as
\begin{eqnarray}
 P_{pol}^{\mu\nu} \Delta
H_{\mu\nu} =  A(u,\rho)+ B(u,\rho)\, \hat p_{{\rm out}}^2  \, ,
\end{eqnarray}
the phase space integration can be splitted into one part that is identical to
the unpolarized case and another one which is purely $d-4$ dimensional. The
results coming from the latter are singled out, writing them under hats, as
they contain soft contributions which have to be factorized.
\begin{eqnarray}
 P_{pol}^{\mu\nu} \Delta
H_{\mu\nu} dPS^{(2)}=  \frac{1}{8\pi}
\frac{(4\pi)^\epsilon}{\Gamma(1-\epsilon)}\, \frac{d\rho}{(1-a(u))}
\left(\frac{Q^2 (1-u)}{u}\right)^{-\epsilon} \left(\frac{(1-\rho)\, (\rho
-a(u))}{(1-a(u))^2} \right)^{-\epsilon} \nonumber\\ \left[ A(u,\rho)
-\frac{\epsilon}{1-\epsilon} \frac{(1-\rho)\, (\rho -a(u))}{(1-a(u))^2}\,
\frac{Q^2 (1-u)}{u}\, B(u,\rho)\right]
\end{eqnarray}
As it has been shown in reference \cite{grau}, the distinctive value
\begin{equation}
\rho =a(u)\equiv \frac{x(1-u)}{u(1-x)}
\end{equation}
represents the configuration where the hadrons are produced in the direction of
the
incoming nucleon, thus giving rise to additional collinear singularities which
do not show up neither in totally inclusive deep inelastic scattering nor in
electron-proton annihilation. For fracture like configurations
it is convenient to use the variable
\begin{equation}
\omega \equiv \frac{1-\rho}{1-a(u)}
\end{equation}
which transforms equation (\ref{dps} ) into the usual two particle phase space
for HVBM
\cite{vog}.

Adding up contributions, we finally find
\begin{eqnarray}
\frac{d \Delta \sigma}{dx\,dy\,dz} & = &
  Y^p \lambda \sum_{i=q,\bar q} c_i \left\{ \int \int_{A} \frac{du}{u}
\frac{d\rho}{\rho}
\left\{
 \Delta q_i(\frac{x}{u})\, D_{q_i}(\frac{z}{\r})\,
\delta(1-u)\delta(1-\r)\frac{}{}  \right.\right.  \nonumber \\
&+& \left.\left. \Delta q_i(\frac{x}{u})\, D_{q_i}(\frac{z}{\r})\,
\frac{\alpha_s}{2\pi} \left[ \p  \left(P_{q\leftarrow q}(\r)\delta(1-u)+
\Delta P_{q\leftarrow q}(u)\delta(1-\r)\right) +C_f \Delta
\Phi_{qq}(u,\r)\right] \right.\right. \nonumber\\
&+& \left.   \left. \Delta q_i(\frac{x}{u})\, D_g(\frac{z}{\r})\,
\frac{\alpha_s}{2\pi} \left[ \p  \left(P_{g\leftarrow q}(\r)\delta(1-u)+
\Delta \hat P_{gq\leftarrow q}(u)\delta(\r-a)\right) +C_f \Delta
\Phi_{qg}^A(u,\r)\right]
 \right. \right.\nonumber\\
&+&  \left.  \left. \Delta g(\frac{x}{u})\, D_{q_i}(\frac{z}{\r})\,
\frac{\alpha_s}{2\pi} \left[ \p  \left(\Delta P_{q\leftarrow g}(u)\delta(1-\r)+
\Delta \hat P_{q\bar{q}\leftarrow g}(u)\delta(\r-a)\right) + T_f \Delta
\Phi_{gq}(u,\r)\right] \right\}\right.  \nonumber\\
&+& \left.  \int_{B} \frac{du}{u} (1-x)
\left\{ \Delta M_{q_i}(\frac{x}{u},(1-x) z)  \left(
\delta(1-u) \frac{}{} + \frac{\alpha_s}{2\pi} \left[ \p  \Delta P_{q\leftarrow
q}(u)+C_f \Delta \Phi_{q}(u,\r) \right] \right) \right. \right.\nonumber \\
&+& \left.\left. \Delta M_g (\frac{x}{u},(1-x) z) \,
 \frac{\alpha_s}{2\pi} \left[ \p \Delta P_{q\leftarrow g}(u)+T_f \Delta
\Phi_g (u,\r)\right] \right\}\right\}
\end{eqnarray}
where
\begin{equation}
{1\over \hat{\epsilon}}
\equiv {1\over \epsilon}
{{\Gamma[1-\epsilon]}\over{\Gamma[1-2
\epsilon]}}\left({{4\pi\mu^2}\over{Q^2}}\right)^\epsilon
=\frac{1}{\epsilon}-\gamma_E+\log(4\pi)+\log(\frac{\mu^2}{Q^2})+\cal{O}(\epsilon)
\end{equation}

The integration ranges for both convolutions, labelled $A$ and $B$, come from
the definition of the variables and momentum conservation and can be found
in appendix B.
The poles proportional to $\delta (1-u)$ correspond to final state
singularities, so are multiplied by unpolarized Altarelli Parisi kernels
$P_{i\leftarrow j}(\rho)$ \cite{ap}. Those proportional to $\delta (1-\rho)$,
are
related to the initial state singularities and are multiplied by spin
dependent kernels $\Delta P_{i\leftarrow j}(u)$ \cite{ap}. The poles
proportional to $\delta (\rho-a)$ are the collinear divergences mentioned
previously and are multiplied by unsubstracted polarized splitting functions
$\Delta \hat P_{ij\leftarrow k}(u)$ \cite{kon}.

The functions $\Delta \Phi (u,\rho)$ are the finite next to leading order
contributions to the cross section. $\Delta \Phi_{i\, j} (u,\rho)$ is a
$i$ ($i=$ quark, gluon) initiated contribution where an outgoing parton
$j$ undergoes hadronization.  $\Delta \Phi_{q\,(g)} (u,\rho)$ are the
quark (gluon) initiated corrections to the fracture processes, which
are identical to those of the totally inclusive polarized structure function.
Notice that  $\Delta \Phi_{q\,g} (u,\rho)$ depends explicitely
on the integration subinterval and the others contain $()_+$ prescriptions and
$\delta$ functions which have support in certain subintervals only.
The expressions for the kernels and the finite NLO terms can be found in
appendix C.\\

\noindent{\large \bf Factorization.}

Having computed the whole cross section up to next to leading order, we are now
able to factorize all the divergences and finite soft terms by means of the
definition of scale dependent distributions.
In order to respect the universal character of these distributions, it is
mandatory to use here the same factorization prescriptions which were
fixed in totally inclusive polarized deep inelastic scattering,
for polarized parton distributions, and in one-particle inclusive
electron-positron annihilation, for fragmentation functions.
Provided this consistency requirement is satisfied, one can adopt any
well defined prescription.

 Fixing the factorization scale equal to $Q^2$, in the $\overline{MS}$ scheme
the prescription ammounts to absorbe only the $1/\hat{\epsilon}$-terms. In the
case of polarized deep inelastic scattering, it has been shown that this scheme
leaves some soft contributions unsubstracted \cite{vog}. Within the HVBM
prescription for
$\gamma_5$ and $\epsilon_{\mu\nu\rho\sigma}$, these contributions can be
identified because they come from the use of helicity projectors for the
initial state partons \cite{vog}. They
are related with the terms coming from de $d-4$ dimensional phase space
integration  (hat terms). It is possible then to define  a slight variation of
the traditional $\overline{MS}$ scheme, called $\overline{MS}_p$
\cite{vog,heavy}, in order to substract the remanent soft contributions. In
general, the definition of the scale dependent quark distributions can be
written as
\begin{eqnarray}
\Delta q_i(\xi) & = &
\int_{\xi}^1 \frac{du}{u} \left\{
\left[ \delta(1-u) + \frac{\alpha_s}{2\pi}\left(
\pp \Delta P_{q\leftarrow q}(u) -  C_{f} \Delta
\tilde{f}_q^F(u) \right)\right]  \Delta q_i(\frac{\xi}{u},Q^2) \right.
\nonumber \\
& & \left.
+  \frac{\alpha_s}{2\pi} \left[
\pp \Delta P_{q\leftarrow g}(u) -  T_{f} \Delta
\tilde{f}_g^F(u) \right]  \Delta g(\frac{\xi}{u},Q^2) \right\}
\end{eqnarray}
In the $\overline{MS}_p$, the finite substraction term $\Delta \tilde{f}_q^F$,
is designed to absorb soft contributions coming from real gluon emission
diagrams and enforces the non-singlet axial current conservation. Conversely,
the term
$\Delta \tilde{f}_g^F$, which absorbes soft contributions coming from
photon-gluon fusion diagrams, leads to the axial anomaly result for the singlet
axial current \cite{strat}. In this way,
the $\overline{MS}_p$ definition of polarized parton distributions guarantees
the conservation of $\Delta \Sigma= \sum_i \int^1_0 \Delta q_i(x,Q^2) dx$,
which implies the scale independence of the net spin carried by quarks.

As there is no need to make finite substractions for {\it unpolarized} final
states, the definition for the scale dependent fragmentation functions is
simply given by
\begin{eqnarray}
 D_{q_i}(\xi) & = &
\int_{\xi}^1 \frac{du}{u} \left\{
\left[ \delta(1-u) + \frac{\alpha_s}{2\pi}
\pp  P_{q\leftarrow q}(u) \right] D_{q_i}(\frac{\xi}{u},Q^2)
 +  \frac{\alpha_s}{2\pi}
\pp P_{g\leftarrow q}(u)\,  D_g(\frac{\xi}{u},Q^2) \right\}
\end{eqnarray}
which is the canonical  $\overline{MS}$ prescription used in ${\rm e}^+\,{\rm
e}^- \rightarrow {\rm h\, X}$.

For polarized fracture functions, the definition of the scale dependent
distributions requires two parts, as in the unpolarized case. One, called
homogeneus, which deals with initial state singularities in fracture like
events, and another one called inhomogeneus, which has to absorb the additional
collinear singularities related to the $\rho=a(u)$ fragmentation
configurations.
In order to be able to substract the finite soft contributions that arise along
the initial state divergences in the HVBM prescription, we also include  the
$\overline{MS}_p$
counterterms $\Delta \tilde{f}^{MH}$ and  $\Delta \tilde{f}^{MI}$, so
\begin{eqnarray}
\Delta M_{q_i}(\xi,\zeta) =
 \int_{\frac{\xi}{1-\zeta}}^1 \frac{du}{u} \left\{
\left[ \delta(1-u) + \frac{\alpha_s}{2\pi} \left(
\pp \Delta P_{q\leftarrow q}(u) -  C_{f} \Delta
\tilde{f}_q^{MH}(u)\right) \right]  \Delta M_{q_i}(\frac{\xi}{u},\zeta,Q^2)
\right. \nonumber \\
\left. + \frac{\alpha_s}{2\pi} \left[
\pp \Delta P_{q\leftarrow g}(u) -  T_{f} \Delta
\tilde{f}_g^{MH}(u) \right]  \Delta M_g(\frac{\xi}{u},\zeta,Q^2) \right\}
\hspace*{38mm}
\nonumber \\
 \hspace*{5mm} +\int_{\xi}^{\frac{\xi}{\xi+\zeta}} \frac{du}{u}
\frac{u}{x(1-u)}
\left\{   \frac{\alpha_s}{2\pi} \left[
\pp \Delta \hat P_{gq\leftarrow q}(u) - C_{f} \Delta
\tilde{f}_q^{MI}(u) \right]  \Delta q_i(\frac{\xi}{u},Q^2)  \,
D_g(\frac{\zeta u}{\xi (1-u)},Q^2) \right. \nonumber \\
 \left. + \frac{\alpha_s}{2\pi} \left[
\pp \Delta \hat P_{q\bar{q}\leftarrow g}(u) - \frac{\alpha_s}{2\pi} T_{f}
\Delta
\tilde{f}_g^{MI}(u) \right]  \Delta g(\frac{\xi}{u},Q^2) \,
D_{q_i}(\frac{\zeta u}{\xi (1-u)},Q^2) \right\}\hspace*{10mm}
\end{eqnarray}
For the homogeneus part, the counterterms $\Delta \tilde{f}^{MH}(u)$ are the
same as those used in polarized inclusive DIS, because the structure of the
corrections is identical.
For the inhomogeneus part, the
counterterms are those associated with the ones found previously for the
homogeneus part, as can be seen in the hat terms of the finite contributions.
This is so because the finite contributions and divergences come
from the same real gluon emission and quark-antiquark gluon splitting,
The substracted cross section can then be written as
\begin{eqnarray}
 \frac{d\Delta \sigma}{dx\,dy\,dz} & = &
  Y^p \lambda \sum_{i=q,\bar q} c_i\left\{\int \int_{A} \frac{du}{u}
\frac{d\r}{\r}
\left\{
\Delta q_i(\frac{x}{u},Q^2)\, D_{q_i}(\frac{z}{\r},Q^2) \,
\delta(1-u)\delta(1-\r)\frac{}{}
\right. \right.  \nonumber \\
&+& \left.\left.
\Delta q_i(\frac{x}{u},Q^2)\, D_{q_i}(\frac{z}{\r},Q^2)  \,
\frac{\alpha_s}{2\pi} C_f \left[
  \Delta \Phi_{qq}(u,\r) - \Delta \tilde{f}_q^F(u,\r) \right] \right.\right.
\nonumber \\
&+&  \left.\left. \Delta q_i(\frac{x}{u},Q^2)\, D_g(\frac{z}{\r},Q^2)\,
\frac{\alpha_s}{2\pi} C_f \left[
 \Delta \Phi_{qg}^A(u,\r) - \Delta \tilde{f}_q^{MI}(u,\r)\right]
 \right. \right.\nonumber\\
& +&   \left. \left. \Delta g(\frac{x}{u},Q^2)\, D_{q_i}(\frac{z}{\r},Q^2)\,
\frac{\alpha_s}{2\pi} T_f\left[
  \Delta \Phi_{gq}(u,\r)- \Delta\tilde{f}_g^F(u,\r) - \Delta
\tilde{f}_g^{MI}(u,\r)\right]
\right\}  \right. \nonumber\\
& +&\left.  \int_{B} \frac{du}{u} (1-x)
\left\{ \Delta M_{q_i}(\frac{x}{u},(1-x) z,Q^2)  \left(
\delta(1-u) \frac{}{} + \frac{\alpha_s}{2\pi} C_f \left[
 \Delta \Phi_{q}(u,\r) - \Delta\tilde{f}_q^{MH}(u) \right] \right)
 \right.\right. \nonumber \\
&+ & \left.\left. \Delta M_g (\frac{x}{u},(1-x) z,Q^2) \,
 \frac{\alpha_s}{2\pi} T_f \left[  \Delta \Phi_g(u,\r)
-\Delta\tilde{f}_g^{MH}(u) \right]
\right\}\right\}
\end{eqnarray}
where the counterterms in the $\overline{MS}_p$ scheme are given by
\begin{eqnarray}
 \Delta\tilde{f}_q^F(u,\r) &=& 4(u-1)\, \delta(1-\rho) \nonumber\\
\Delta \tilde{f}_q^{MI}(u,\r) &=& 4(u-1)\, \delta(\rho -a) \nonumber\\
  \Delta\tilde{f}_q^{MH}(u) &=& 4(u-1)\nonumber\\
 \Delta\tilde{f}_g^F(u,\r) &=& 2(1-u)\, \delta(1-\rho)\nonumber\\
\Delta \tilde{f}_g^{MI}(u,\r) &=& 2(1-u)\, \delta(\rho-a)\nonumber\\
 \Delta\tilde{f}_g^{MH}(u) &=& 2 (1-u)
\end{eqnarray}
in the case of the light quarks (u, d, s) and $0$ for heavy quarks
\cite{heavy}.
Notice that in the $\overline{MS}$ scheme, all of these counterterms are
choosen to be $0$.\\

\noindent{\large \bf Conclusions.}\\

We have calculated the ${\cal{O}}(\alpha_s)$ one-particle inclusive cross
section
in polarized deep inelastic lepton-hadron scattering, showing that with the
inclusion of polarized fracture functions it is possible to consistently
factorize all the collinear singularities that occur and that, within the HVBM
prescription, the  $\overline{MS}_p$ scheme can be straightforwardly applied in
order to factorize unwanted finite soft contributions. In this way, the
$\overline{MS}_p$ scheme guarantees the conservation of the non-singlet axial
current and the non-conservation of the singlet one, as dictated by the anomaly
result. This requirement allows the definition of polarized parton and fracture
distributions intimately related to the fraction of the nucleon spin carried by
partons.

Having defined an universal and physically meaningfull factorization scheme
for both current and target fragmentation, consistent with those used in
totally inclusive spin dependent deep inelastic scattering and unpolarized
electron  proton annihiliation, it will be possible to perform an unanmbiguous
${\cal O}(\alpha_s)$  analysis of forthcoming inclusive experiments.
\\

\noindent{\large \bf Acknowledgements.} \\

We greatly acknowledge L. N. Epele and H. Fanchiotti for fruitful discussions.

\pagebreak

\noindent{\large \bf Appendix A.}\\

The projection of the real gluon emission matrix element, within the HVBM
prescription, is given by
\begin{eqnarray}
 P_{pol}^{\mu\nu} \Delta H_{\mu\nu} &=& -4\pi
{{\alpha_s}\over{2\pi}} Q^2_q 2\pi  {{\alpha^2}\over {S_H x}} C_f
\left({{4\pi\mu^2}\over{Q^2}}\right)^\epsilon {{u}\over {Q^2}}   \\
& \times &
\left[ {{4\,\left( 1 + { \epsilon} \right) \,\left( { s_{ig}} - { s_{iq}}
\right) \,
{s_{qg}}}\over {{ s_{ig}}}} +
{{4\,\left( -1 + { \epsilon} \right) \,{ s_{ig}}\,
     \left( { s_{ig}} + { s_{iq}} \right) }\over {{ s_{qg}}}}
 -  8\,{ \epsilon}\,{ s_{ig}}  \right. \nonumber \\
& - & \left. {{8\,{ s_{iq}}\,
      \left( {{{ s_{ig}}}^2} + {Q^2}\,{ s_{iq}} + { s_{ig}}\,{ s_{iq}} \right)
      }\over {{ s_{ig}}\,{ s_{qg}}}}
 -{{8\,{{\left( { s_{ig}} + { s_{iq}} \right) }^2}\,
     \left( -{ s_{ig}} + { \epsilon}\,{ s_{ig}} - { s_{qg}} -
       { \epsilon}\,{ s_{qg}} \right) \,\hat p_{out}^{2}}\over
   {{{{ s_{ig}}}^2}\,{ s_{qg}}}} \right]\nonumber
\end{eqnarray}
where $s_{AB}=2\,p_A \cdot p_B$. The labels $q$ and $g$ take the values 1 and
2, respectively, for quark fragmentation, or 2 and 1 for gluon fragmentation.

For photon gluon fusion
\begin{eqnarray}
 P_{pol}^{\mu\nu} \Delta
H_{\mu\nu} &=&4\pi {{\alpha_s}\over{2\pi}} Q^2_q 2\pi {{\alpha^2}\over {S_H x}}
T_f \left({{4\pi\mu^2}\over{Q^2}}\right)^\epsilon  \\
& \times & \left[ {{-4\,\left( -2\,{Q^2} + {\it s_{iq}} + {\it s_{i\bar q}}
\right) \,
     \left( {{{\it s_{iq}}}^2} + {{{\it s_{i\bar q}}}^2} \right) \,u}\over
   {{Q^2}\,{\it s_{iq}}\,{\it s_{i\bar q}}}}+
{{8\,{{\left( {\it s_{iq}} + {\it s_{i\bar q}} \right) }^2}\,
     \left( {{{\it s_{iq}}}^2} + {{{\it s_{i\bar q}}}^2} \right) \, u \, \hat
p_{out}^{2}}\over
   {{Q^2}\,{{{\it s_{iq}}}^2}\,{{{\it s_{i\bar q}}}^2}}} \right]  \nonumber
\end{eqnarray}
The expression for $s_{AB}$ in terms of the variables $\rho,u$ and $\omega$ can
be found in reference \cite{grau}. \\

\noindent{\large \bf Appendix B.}\\

The integration range $A$ is splitted into two subintervals
\begin{equation}
A_1\, : \, u \in \left[x\, , \frac{x}{x+(1-x) z}\right] \, , \ \ \rho \in
\left[a(u)\, , 1\right]
\end{equation}
and
\begin{equation}
A_2\, : \, u  \in  \left[\frac{x}{x+(1-x) z}\, , 1\right] \, , \ \ \rho \in
\left[z\, , 1\right]
\end{equation}
while $B$ is given by
\begin{equation}
B \,:\, u \in \left[\frac{x}{x-(1-x) z}\, , 1\right]
\end{equation} \\

\noindent{\large \bf Appendix C.}\\

The splitting functions are given by \cite{ap,kon}
\begin{eqnarray}
\Delta P_{q\leftarrow q}(u)&=&
C_f \left[ 2\left(\frac{1}{1-u}\right)_+ + \frac{3}{2}
\delta(1-u)-1-u\right]
\nonumber \\
\Delta P_{q\leftarrow g}(u)&=&T_f \left[2u-1\right]\nonumber \\
\Delta \hat P_{gq\leftarrow q}(u)&=&
C_f \left[ \frac{1+u^2}{1-u}\right]\nonumber \\
 P_{q\leftarrow q}(u)&=&
C_f \left[ 2\left(\frac{1}{1-u}\right)_+ + \frac{3}{2}
\delta(1-u)-1-u\right]
\nonumber \\
P_{g\leftarrow q}(u)&=&C_f \left[ 2\frac{1}{u} -2+u \right]
\nonumber \\
\Delta \hat{P}_{q\bar{q}\leftarrow g}(u)&=&T_f \left[2u-1\right]
\end{eqnarray}
The finite next to leading order contributions are
\begin{eqnarray}
& & \Delta\Phi_{qq}(u,\rho) = \nonumber\\
& - & 8\, \delta(1-r) \delta(1-u) + \left[(1-r)+\frac{1+r^2}{1-r}
\log(r)-(1+r)\log(1-r)+2\left(\frac{\log(1-r)}{(1-r)} \right)_{+}\right]
\delta(1-u)\nonumber\\
&+& \left[-(1-u)+\frac{1+u^2}{1-u}
\log(\frac{1-x}{u-x})-(1+u)\log(1-u)+2\left(\frac{\log(1-u)}{(1-u)} \right)_{+}
-2\widehat{(1-u)}\right]
\delta(1-r) \nonumber\\
&+& 2 \left(\frac{1}{1-r}\right)_+ \left(\frac{1}{1-u}\right)_+  -
\left(\frac{1}{1-r}\right)_+ (1+u)- \left(\frac{1}{1-u}\right)_+(1+r)
 -  {{2\,\left( 1 - u \right) \,u\,\left( 1 - x \right) }\over {u -
x}}\nonumber\\
&-&{{\left( 1 - r \right) \,\left( 1 - u +
        \left( 1 - x \right) \,\left( -1 +
           u\,\left( 1 + 2\,u \right) \,\left( 1 - x \right)  - x \right)
 \right) }\over {{{\left( u - x \right) }^2}}} +
  {{4\,u\,\left( 1 - x \right) }\over {u - x}}-  {{2\,x}\over {u - x}}
\end{eqnarray}
\begin{eqnarray}
\Delta \Phi_{qg}^{A=1}(u,\rho) & = & \delta (\r-a)
\left[-2\widehat{(1-u)}-(1-u) +
{{1+u^2}\over{1-u}} \log \left( {{(1-x)(1-u)} \over {(u-x) }}\right) \right]
\nonumber \\
& &  +\left({1\over{\r-a}}\right)_+  {{1+u^2}\over{1-u}}
 - {{\r\,{u^2}\,{{\left( 1 - x \right) }^2}}\over
    {{{\left( u - x \right) }^2}}} +
  {{\r\,{u^3}\,{{\left( 1 - x \right) }^2}}\over
    {\left( 1 - u \right) \,{{\left( u - x \right) }^2}}}  \nonumber \\
& & -  {{2\,{u^3}\,\left( 1 - x \right) }\over
    {\left( 1 - u \right) \,\left( u - x \right) }} +
  {{u\,\left( 1 - x \right) \,x}\over {{{\left( u - x \right) }^2}}} -
  {{2\,{u^2}\,\left( 1 - x \right) \,x}\over {{{\left( u - x \right) }^2}}}
\end{eqnarray}
\begin{eqnarray}
\Delta \Phi_{qg}^{A=2}(u,\rho) & = & \delta (1-u) \left[\r +  \left(\r+{2\over
\r}-2 \right)
\log \left(\r(1-\r) \right) \right]
+\left({1\over{1-u}}\right)_+ \left(\r+{2\over \r}-2 \right)  \nonumber\\
& & - {{2\,{{\left( 1 - \r \right) }^2}}\over {\r\,\left( 1 - u \right) }} +
  {{1 - u}\over { r-a}} - {{2\,\left( 1 - u \right) \,u\,
      \left( 1 - x \right) }\over {u - x}} +
  {{2\,{{\left( 1 - \r \right) }^2}\,{u^3}\,{{\left( 1 - x \right) }^2}}\over
      {\left(  \r-a \right) \,\left( 1 - u \right) \,
      {{\left( u - x \right) }^2}}}  \nonumber\\
& &  +{{\left( 1 - \r \right) \,u\,\left( 1 - x \right) \,x}\over
    {{{\left( u - x \right) }^2}}} + {{\left( 2 - \r \right) \,x}\over {u - x}}
\end{eqnarray}
\begin{eqnarray}
\Delta\Phi_{gq}(u,\rho) &=& (\delta (1-\r)+\delta (\r-a))
\left[2\widehat{(1-u)} + (2 u-1) \log \left(
{{(1-x)(1-u)} \over {(u-x) }} \right) \right]  \nonumber \\
& & +(2 u-1) \left[ \left({1\over{1-\r}}\right)_+
+\left({1\over{\r-a}}\right)_+ -2 u{{1-x}\over{u-x}} \right]
\end{eqnarray}
\begin{eqnarray}
\Delta \Phi_{q}(u,\rho) & = &
-\frac{1+u^2}{1-u}\log(u)-(1+u)\log(1-u)
+2\left(\frac{\log(1-u)}{(1-u)} \right)_{+}-\frac{3}{2}
 \left(\frac{1}{1-u}\right)_+ \nonumber \\
& & + 3 u + \frac{7}{2}\delta(1-u) -2\widehat{(1-u)}
\end{eqnarray}
\begin{eqnarray}
\Delta \Phi_{g} (u,\rho)& = & 2\widehat{(1-u)} +(2 u-1) \log \left(
{{(1-u)}\over u}-1  \right)
\end{eqnarray}

\pagebreak

\pagebreak

\noindent{\large \bf Figure Captions}
\\

\begin{enumerate}
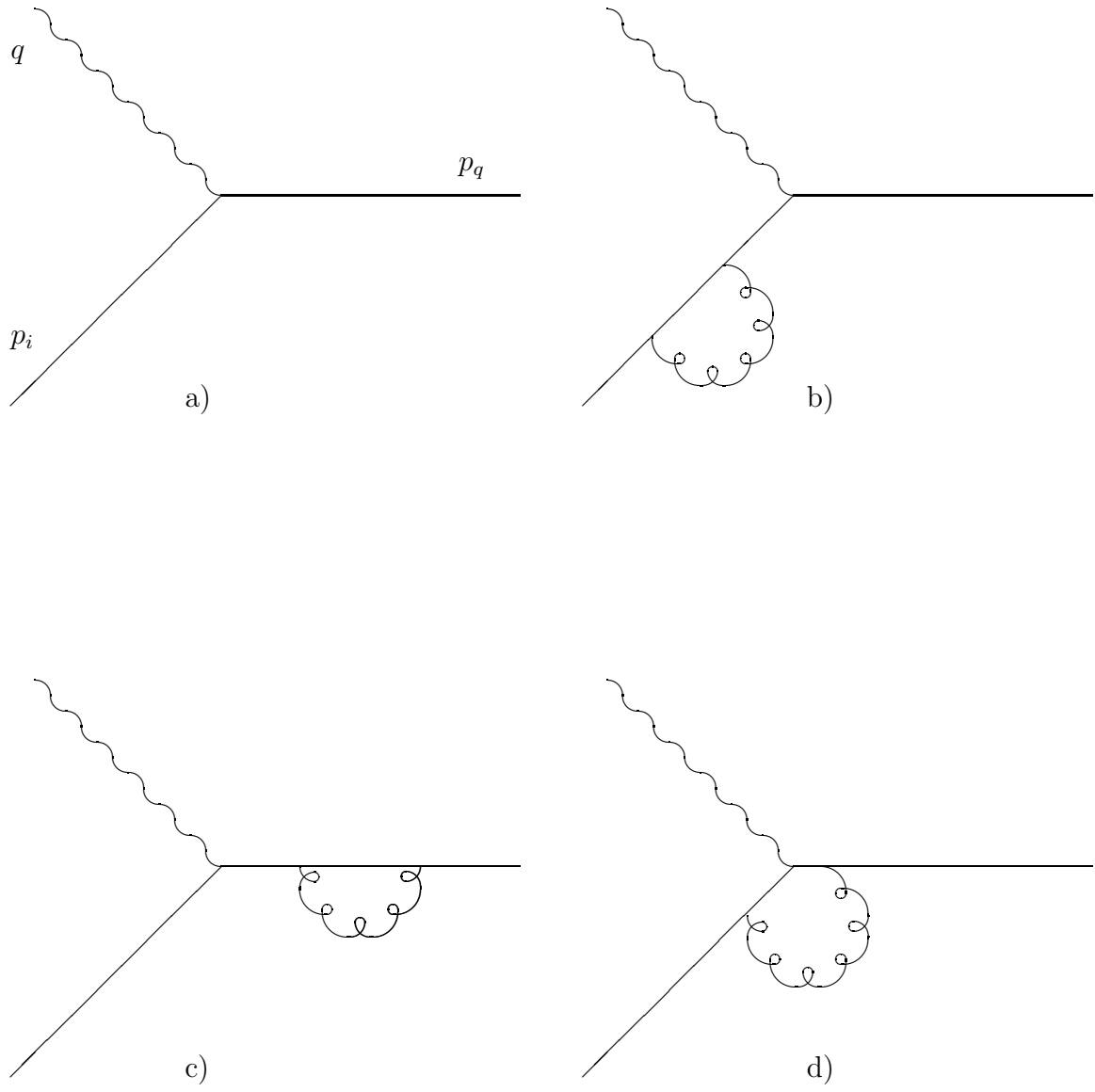
\item[Figure 1 ] a) Lowest order parton-photon graph ; b),c) and d) virtual
gluon correction graphs to a).
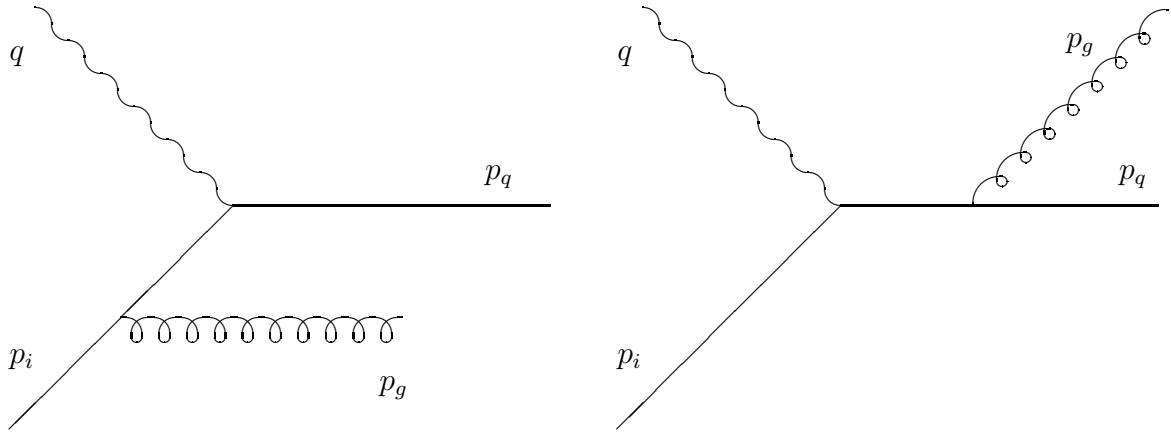
\item[Figure 2 ] Real gluon emission corrections to 1a).
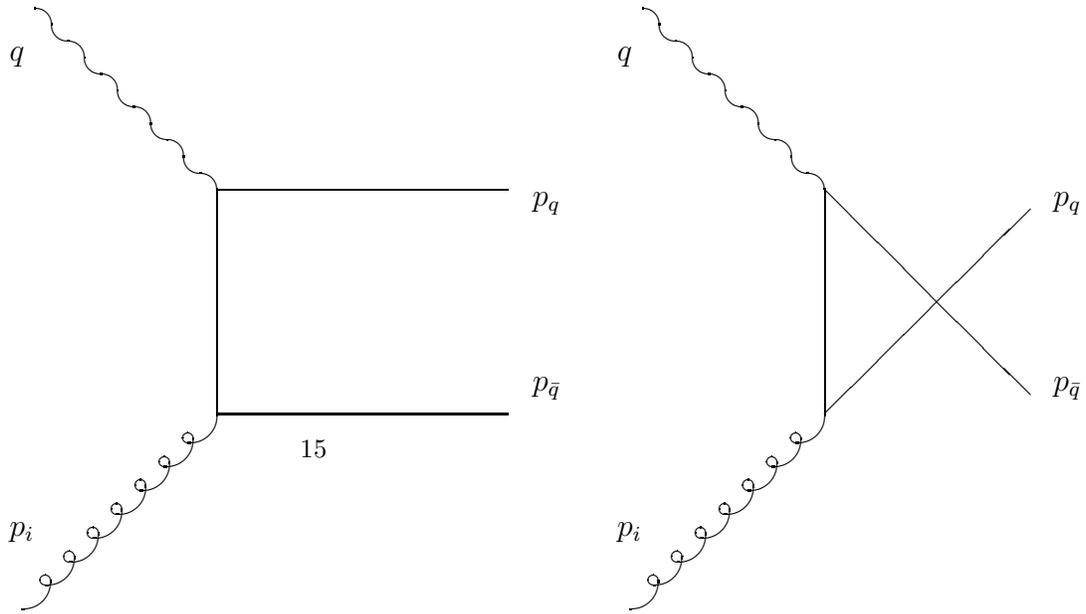
\item[Figure 3 ] Gluon contribution  at order $\alpha_{s}$
\end{enumerate}

\pagebreak
\input FEYNMAN
\textheight 410pt \textwidth 200pt
\begin{picture}(10000,18000)
\drawline\photon[\SE\REG](-2000,18000)[12]
\put(-3000,16000){\large  $q$}
\drawline\fermion[\E\REG](\photonbackx,\photonbacky)[12000]
\put(-3000,4500){\large $p_i$}
\drawline\fermion[\SW\REG](\photonbackx,\photonbacky)[12000]
\put(15000,11500){\large $p_q$}
\put(4000,2000){\large a)}
\drawline\photon[\SE\REG](21000,18000)[12]
\drawline\fermion[\E\REG](\photonbackx,\photonbacky)[12000]
\drawline\fermion[\SW\REG](\photonbackx,\photonbacky)[4000]
\drawloop\gluon[\SE\FLIPPEDFLAT](\fermionbackx,\fermionbacky)
\drawline\fermion[\SW\REG](\fermionbackx,\fermionbacky)[8000]
\put(29000,2000){\large b)}

\drawline\photon[\SE\REG](-2000,-9000)[12]
\drawline\fermion[\E\REG](\photonbackx,\photonbacky)[8000]
\drawloop\gluon[\S\FLIPPEDFLAT](\fermionbackx,\fermionbacky)
\drawloop\gluon[\S\FLIPPEDCURLY](\fermionbackx,\fermionbacky)
\drawline\fermion[\E\REG](\fermionbackx,\fermionbacky)[4000]
\drawline\fermion[\SW\REG](\photonbackx,\photonbacky)[12000]
\put(4000,-25000){\large c)}

\drawline\photon[\SE\REG](21000,-9000)[12]
\drawline\fermion[\E\REG](\photonbackx,\photonbacky)[1000]
\drawloop\gluon[\SE\CENTRAL](\fermionbackx,\fermionbacky)
\drawline\fermion[\E\REG](\fermionbackx,\fermionbacky)[11000]
\drawline\fermion[\SW\REG](\photonbackx,\photonbacky)[12000]
\put(29000,-25000){\large d)}

\end{picture}

\begin{center}
\vspace*{100mm}
{\large \bf Figure 1}
\end{center}

\newpage
\begin{picture}(10000,18000)
\drawline\photon[\SE\REG](-2000,18000)[12]
\put(-3000,16000){\large  $q$}
\drawline\fermion[\E\REG](\photonbackx,\photonbacky)[12000]
\drawline\fermion[\SW\REG](\photonbackx,\photonbacky)[6000]
\drawline\gluon[\E\REG](\fermionbackx,\fermionbacky)[10]
\drawline\fermion[\SW\REG](\fermionbackx,\fermionbacky)[6000]
\put(-3000,4500){\large $p_i$}
\put(15000,11500){\large $p_q$}
\put(11000,3500){\large $p_g$}

\drawline\photon[\SE\REG](21000,18000)[12]
\drawline\fermion[\SW\REG](\photonbackx,\photonbacky)[12000]
\drawline\fermion[\E\REG](\photonbackx,\photonbacky)[5000]
\drawline\gluon[\NE\REG](\fermionbackx,\fermionbacky)[7]
\drawline\fermion[\E\REG](\fermionbackx,\fermionbacky)[7000]
\put(20000,16000){\large  $q$}
\put(20000,4500){\large $p_i$}
\put(39000,11500){\large $p_q$}
\put(37000,16500){\large $p_g$}

\drawline\photon[\SE\REG](-2000,-9000)[11]
\drawline\fermion[\S\REG](\photonbackx,\photonbacky)[8500]
\drawline\gluon[\SW\REG](\fermionbackx,\fermionbacky)[7]
\drawline\fermion[\E\REG](\photonbackx,\photonbacky)[11000]
\drawline\fermion[\E\REG](\gluonfrontx,\gluonfronty)[11000]
\put(-3000,-29000){\large  $p_i$}
\put(-3000,-11000){\large  $q$}
\put(16800,-16500){\large  $p_q$}
\put(16800,-23500){\large  $p_{\bar q}$}

\drawline\photon[\SE\REG](21000,-9000)[11]
\drawline\fermion[\S\REG](\photonbackx,\photonbacky)[8500]
\drawline\gluon[\SW\REG](\fermionbackx,\fermionbacky)[7]
\drawline\fermion[\SE\REG](\photonbackx,\photonbacky)[11000]
\drawline\fermion[\NE\REG](\gluonfrontx,\gluonfronty)[11000]
\put(20000,-29000){\large  $p_i$}
\put(20000,-11000){\large  $q$}
\put(36500,-16500){\large  $p_q$}
\put(36500,-23500){\large  $p_{\bar q}$}
\put(19500,-36500){\large \bf Figure 3 }

\end{picture}
\begin{center}
{\large \bf Figure 2 }
\end{center}
\end{document}